\documentclass[Journal,letterpaper,InsideFigs,NoLineNumbers]{ascelike-new}
\usepackage[utf8]{inputenc}
\usepackage[T1]{fontenc}
\usepackage{booktabs}
\usepackage{lmodern}
\usepackage{graphicx}
\usepackage[figurename=Fig.,labelfont=bf,labelsep=period]{caption}
\usepackage{subcaption}
\usepackage[per-mode=symbol]{siunitx}
\usepackage{amsmath}
\usepackage{amsfonts}
\usepackage{amssymb}
\usepackage{amsbsy}
\usepackage{newtxtext,newtxmath}
\usepackage[colorlinks=true,citecolor=black,linkcolor=black]{hyperref}
\usepackage{cleveref}
\begin{document}

\title{Conductive and convective heat transfer in inductive heating of subsea buried pipelines}

\author[1]{Krishna Kumar}
\author[2]{Chadi El Mohtar}
\author[2]{Robert Gilbert}

\affil[1]{Civil, Architectural and Environmental Engineering, University of Texas at Austin. Email: krishnak@utexas.edu}
\affil[2]{Civil, Architectural and Environmental Engineering, University of Texas at Austin}

\maketitle

\begin{abstract}
Inductive heating with high-voltage cables reduces the risk of hydrate formation by raising the temperature of the production fluid in pipelines. Heating the pipeline results in losing a certain fraction of the heat to the surrounding soil through conduction or convection-dominated flow through the soil. However, the amount of heat lost in conduction versus convection and the transition from conduction to convection-dominated heat loss remains unknown. Soil permeability, temperature gradient between cable and mudline, and burial depth influence the mode of heat transfer and the amount of heat lost. We study the dominant mode of heat transfer in pipelines with inductive heating using 2D Finite Difference analysis under different soil and environmental conditions. Low permeability soils primarily exhibit conductive heat transfer, thus losing minimum heat to the surrounding soil. In contrast, convective flow drives a significant fraction of the heat away from the pipeline and towards the ground surface for highly permeable soils, barely heating the fluid in the pipe.  We identify a critical Rayleigh-Darcy number of 1 as the controlling value separating conduction and convection-dominated heat transfer. An increase in burial depth deteriorates the heating efficiency in convection-dominated high permeability soils, while it remains unaffected in conduction-dominated low permeability soils.
\end{abstract}

\section{Background}
Hydrate formation in subsea oil and gas production systems undermines the safety and transport of production fluids. Hydrates form at temperatures below \SI{25}{\celsius} as the unprocessed production fluid cools down~\cite{nazeri2014evaluation,daraboina2015natural} causing pipeline blockage. Deepwater pipelines are particularly vulnerable to hydrate formation since the surrounding water near the seabed is constantly cold. Hence, it is essential to maintain the well-stream above a specific temperature to achieve steady flow and prevent unfavorable hydrate formation. Traditionally, chemicals are injected into the production fluid to lower the critical temperature below which hydrate formation occurs. An alternative to chemical processing is the Direct Electrical Heating (DEH) of the pipelines by forcing a single-phase high electric current through the pipeline itself~\cite{nysveen2005direct}. Unlike chemical injection, electrical heating offers uniform treatment along the entire length of long pipelines. However, DEH is inefficient, sacrificing a  significant fraction of heat to the surrounding seawater while increasing the risk of stray current and pipeline corrosion.

Inductive heating is an alternative to DEH, which involves heating the pipeline indirectly through separate high-voltage (HV) piggyback power cables strapped to the pipeline (see~\cref{fig:model}). The HV current-carrying power cables generate heat through ohmic loss, which then inductively heat the pipeline. In inductive heating, the distance between the power cables (heating elements) and the pipeline wall needs to be small to avoid excessive heat dissipation into the seawater. A significant advantage of the indirect heating method is the galvanic insulation between the electric power cable and the pipeline. The risk of stray currents and corrosion is absent~\cite{nysveen2005direct}. Inductive heating offers easy installation avoiding complex designs to embed cables inside the thermal insulation. 

As the cable temperature rises to inductively heat the pipeline, some heat is lost to the surrounding soil and eventually to the seawater. The HV cables require higher reactive power to compensate for the heat dissipation into the seawater. The dissipated power depends on the electrical and magnetic properties of the pipeline and the cable, and the hydraulic and mechanical properties of the surrounding soil.

Based on the thermal and hydraulic characteristics of the soil, conduction or convection mechanism can dominate the heat transfer process. Conduction is the transfer of heat through thermal resistance of the material without any bulk motion of matter. On the other hand, buoyancy-driven convection transfers heat through fluid flow due to a density difference from the thermal gradient. As the fluid near the cables absorbs the heat, it becomes less dense and rises due to thermal expansion. The surrounding cooler fluid then moves to replace it forming a natural/free convection. Convective flow results in excessive heat loss and is undesirable. Various factors such as temperature gradient, burial depth, soil permeability, soil density, fluid viscosity, and thermal characteristics of the porous media control the mode of heat transfer, i.e., conduction or convection. 

The amount of heat generated in the cables and the ability to dissipate the heat to the surrounding soil determines the sizing of the HV cables. At present, the maximum operating conductor temperature of HV cables is \SI{90}{\celsius} which can translate to cable surface temperatures of up to \SI{70}{\celsius}~\cite{hughes2015effect,swaffield2008methods}. The current industry standard for determining the current rating of cables relies on IEC 60287 or the Neher and McGrath (1957) equation~\cite{neher1994calculation,IEC60287:2006}. The sizing equations consider the electrical load characteristics and thermal conductivity of different sections of the cable materials. Both methods incorrectly assume the heat transfer occurs only through the thermal resistance in the soil (conduction), completely ignoring the convective heat dissipation.

An experimental study by~\citeN{emeana2016thermal} and numerical modeling by~\citeN{hughes2015effect} and~\citeN{t2019combined} show that the current approach to cable sizing underestimates the amount of heat transfer through cables by only considering conductive heat transfer through the soil.~\citeN{hughes2015effect} simulate the submarine environment as a porous media through which seawater can flow, allowing for an increased heat transfer through natural convection from the cables to the seabed. They identify permeability as the critical factor determining the mode of heat transfer. At low permeabilities, conductive heat transfer is dominant with a characteristic uniform symmetric distribution of heat around the cable. As the permeability increases, convection drives more heat towards the ground surface, breaking the symmetric heat distribution around the cable. An intermediate situation also occurs where both conduction and convection contribute to the total heat transfer. 

Besides soil permeability, an increase in the cable temperature or the burial depth causes an earlier onset of convective flow indicating other factors also influence the transition from conduction to convection. The mode of heat transfer depends on many factors: soil parameters (thermal conductivity, permeability) and boundary conditions (cable geometry, burial depth, flux/thermal boundaries). It is unclear when convection becomes the dominant mode of heat transfer and the amount of heat dissipated in conduction and convection. 

Experimental results on heat transfer through saturated glass beads show that high permeable soil increases the convective heat transfer~\cite{fand1986natural}. \citeN{hardee1976boundary} and~\citeN{merkin1979free} developed analytical solutions for heat transfer from a cylinder embedded in a porous media considering natural convective flow. However, none of these authors identified a conduction region.~\citeN{nield1968onset}  separated conduction and convection regions using a non-dimensional Rayleigh-Darcy number.~\citeN{nield1968onset} established a critical Rayleigh-Darcy number of 12 separating conduction from convection-dominated flows for a rectangular domain with a constant heat source. The applicability of this critical Rayleigh-Darcy number of 12 to this submarine condition is unknown due to differences in boundary conditions and asymmetric nature of the problem. 

The Rayleigh number only quantifies the proportion of conduction or convection flow, and not the amount of heat transfer through these different modes. Although the heat transfer is related to the type of flow in the porous media, the relationship is not linear. Hence, we need to identify the relationship between the amount of conductive and convective flow and the corresponding heat loss in each mode. By computing the amount of heat lost in convection and conduction, we will evaluate the favorable and unfavorable conditions for inductive heat transfer in heating submarine pipelines.

In this paper, we numerically model a pipeline heated through inductive transfer from three HV piggyback cables in different soil conditions to determine the different modes of heat transfer. We analyze the heat distribution in the system by varying the soil permeability, cable temperature, and burial depth to quantify the amount of heat loss in different modes of heat transfer. The next section introduces the Rayleigh-Darcy number for classifying conduction and convection-dominated flows. 

\section{Rayleigh-Darcy number}
The non-dimensional Rayleigh number ($Ra^\prime$) determines the flow behavior of fluid (conduction or convection-dominated flows) in a non-uniform mass density field. The Rayleigh number is the ratio of the time scale for thermal transport through conduction (diffusion) to the time scale for convective thermal transport at velocity $u$:

\begin{equation}
Ra^\prime = \frac{\text{time scale for thermal transport via diffusion}}{\text{time scale for thermal transport via convection at velocity }u} = \frac{D^2/\alpha}{\mu/\Delta \rho D g} \,,
\label{eq:ra-text}
\end{equation}
\noindent where $D$ is the depth of burial (or characteristic length), $\alpha$ is thermal diffusivity (\si{\meter\squared\per\second}), $\mu$ is the dynamic viscosity of fluid (\si{\pascal\cdot\second}), $\Delta \rho$ is the mass density difference due to heating (\si{\kilogram\per\meter\cubed}), and $g$ is the acceleration due to gravity (\si{\meter\per\second\squared}). The time scale for conduction across a distance $D$ is $D^2/\alpha$ and the convective velocity $u \sim \Delta \rho D^2 g / \mu$. When the Rayleigh number, $Ra^\prime$, is below a critical value, there is no flow, and heat transfer is purely by conduction; when $Ra^\prime$ exceeds the threshold value, heat transfer is by natural convection. The critical $Ra_c^\prime$ that determines the transition from conduction to convection depends on the geometry (thermal and hydraulic boundary conditions) and type of heat source (constant temperature or flux). 

For a flow through porous media, the Rayleigh number ($Ra^\prime$) is modified accounting for seepage velocity using Darcy's law. The modified Rayleigh-Darcy number ($Ra$) is:

\begin{equation}
Ra = \frac{\rho g D^3 \beta \Delta T}{\mu \alpha} \cdot \frac{k}{D^2} = \frac{\rho g D k \beta \Delta T}{\mu \alpha}\,,
\label{eq:ra}
\end{equation}
where $k$ is the intrinsic permeability (\si{\meter\squared}) and $\beta$ is thermal expansion of water (\si{\per\kelvin}). 

\section{Numerical Model}
We use a 2D Finite Difference model to solve the inductive heat transfer from the HV piggyback cables to the pipeline and eventually to the surrounding soil. We consider both conductive and convective heat transfer mechanisms by solving the partial differential equations for the heat transfer and the time-independent coupled fluid flow. 

\subsection*{Governing Equations} 
The heat transfer in the presence of a constant source $Q_{in}$ under steady-state is: 
\begin{equation}
Q_{in} = - \lambda \nabla^2 T + \rho_f c_{p_f} \mathbf{u} \cdot \nabla T\,,
\label{eq:pde-heat}
\end{equation}
where $\lambda$ is the thermal conductivity (\si{\watt\per\meter-\kelvin}), $T$ is Temperature (\si{\celsius}), $\rho_f$ is the density of fluid (\si{\kilo\gram\per\meter\cubed}), $c_{p_f}$ is the specific heat capacity of fluid (\si{\joule\per\kilo\gram\per\celsius}), and $\mathbf{u}$ is the fluid velocity (\si{\meter\per\second}). The right hand side of~\cref{eq:pde-heat} represents the conductive ($- \lambda \nabla^2 T$) and convective ($ \rho_f c_{p_f} \mathbf{u} \cdot \nabla T$) heat transfer mechanisms. The bulk thermal conductivity of the soil $\lambda$ controls the conductive heat transfer. The bulk thermal conductivity is a function of porosity ($n$) and the individual thermal conductivities of solid ($\lambda_s$) and fluid ($\lambda_f$):

\begin{equation}
\lambda = \lambda_s ( 1 - n) + \lambda_f n\,.
\end{equation}

We assume the soil is fully saturated, i.e., the voids in the soil are filled with water. Darcy's law describes the fluid flow through the porous media resulting from density gradient, $\rho = \rho_0 (1 - \beta \Delta T)$, caused by the thermal differences:

\begin{equation}
u = - \frac{1}{n \mu} k \left(\nabla p + g \rho_{f_0} \left(1 - \beta (T - T_0)\right)\right)\,,
\end{equation}
where $u$ is the fluid flow velocity (\si{\meter\per\second}), $n$ is the porosity, $\mu$ dynamic viscosity (\si{\pascal\cdot\second}), $k$ is the intrinsic permeability (\si{\meter\squared}), $p$ is pressure (\si{\pascal}), $\rho_{f_0}$ is the reference fluid density at ambient temperature, and $\beta$ is the volumetric coefficient of thermal expansion (\si{\per\kelvin}). 

We solve the governing equation using a forward-difference in time, central-difference in space, and an upwind scheme for solving the derivatives in the flow field. 

\subsection*{Geometry and Mesh}

\begin{figure}[htbp]
    \centering
    \includegraphics[width=0.7\linewidth]{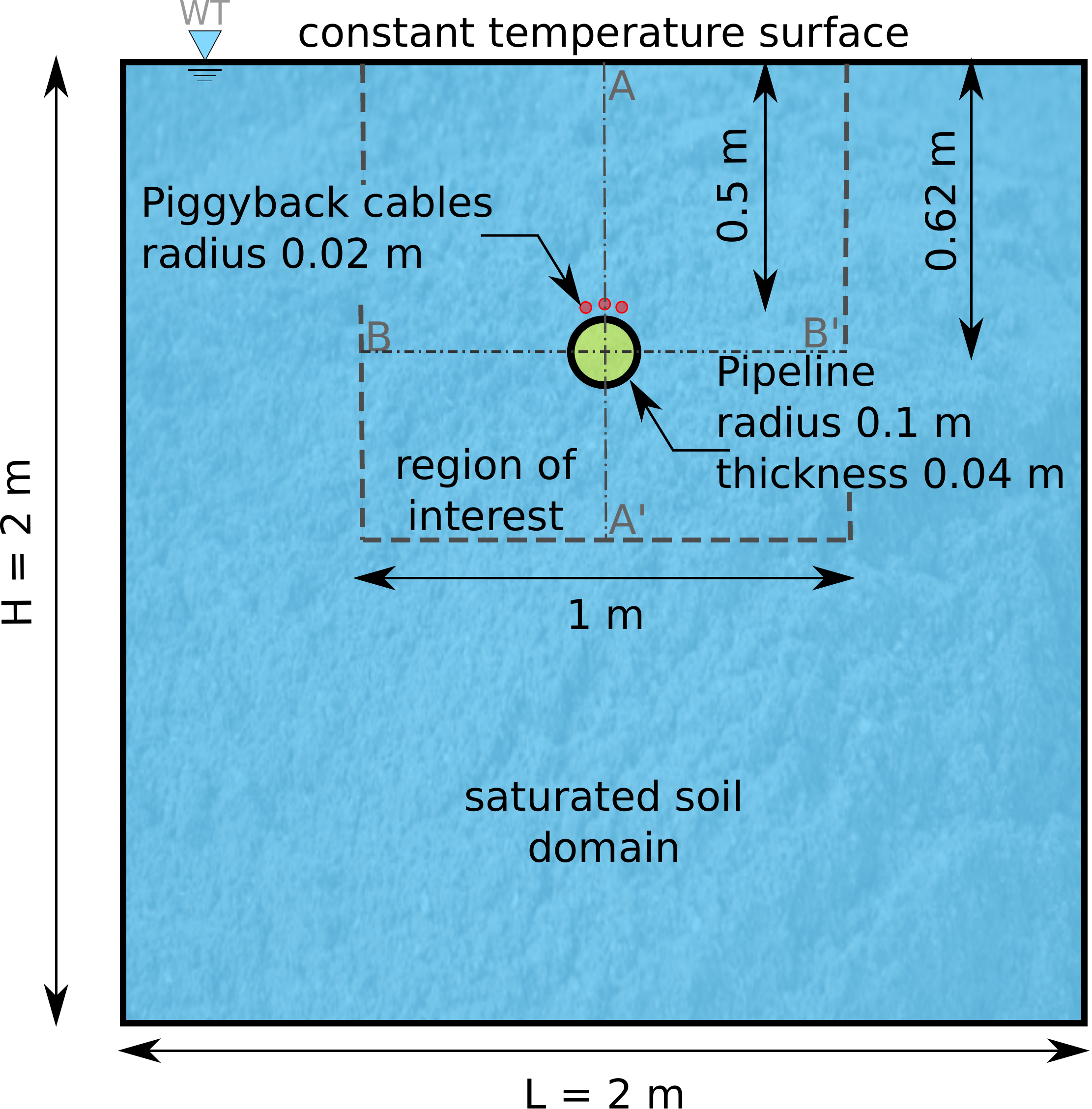}
    \caption{Schematic of the simulation geometry (not to scale). A fine mesh size of 0.004 m was used to discretize the domain with 250,000 elements.}
    \label{fig:model}
\end{figure}

We evaluate the thermal distribution around an oil pipeline with an external diameter of \SI{0.2}{\meter} a wall thickness of \SI{0.04}{\meter}. The oil pipeline is heated by three piggyback cables (diameter: \SI{0.02}{\meter}) at its crown. The piggyback cables are buried at \SI{0.5}{\meter} from the ground surface. We simulate a \SI{2}{\meter} x \SI{2}{\meter} saturated soil domain using the Finite-Difference approach with a very fine mesh size of \SI{4}{\milli\meter}.~\Cref{fig:model} shows the model setup. The upwind scheme is stable when the Courant–Friedrichs–Lewy condition (CFL) is satisfied for a problem with a mesh size of $\Delta x$ and a flow velocity of $u$:

\begin{equation}
c = |\frac{u \Delta t_c}{\Delta x}| \le 1\,.
\end{equation}
\noindent We use a timestep of $\Delta t = \Delta t_c / 100$, where $\Delta t_c$ is the critical time step from the CFL criteria. The domain is sufficiently large to avoid any boundary interference and has no observable influence on the temperature and velocity field variables directly surrounding the cable. The bottom and side boundaries have no fluid flow and no heat flux conditions:

\begin{equation}
n \cdot u = 0 \text{ and } n \cdot (-\lambda \nabla T) = 0\,.
\end{equation}

We maintain a constant temperature at the top boundary to simulate the seawater. The interface between different materials (such as cable/pipeline and soil) is a no-fluid flow boundary but allows heat transfer through the interface. We apply a constant temperature on the three HV piggyback cables to heat the fluid inside the pipeline. All temperatures reported in the paper are temperatures above ambient conditions. We are interested in the heat distribution in a \SI{1}{\meter} x \SI{1}{\meter} subdomain of the problem encompassing the cables and the pipeline as shown in~\cref{fig:model}.

\subsection*{Material Properties}
We model the heat dissipation mechanisms for pipelines buried in different soil types from coarse-grained gravels and sands to fine-grained silts and clay. Soil permeability can vary over several orders of magnitude based on the porosity, mean grain radius, and grain size distribution. The semi-empirical Kozeny-Carman equation relates permeability to porosity and the mean grain size $d_m$:

\begin{equation}
k = \frac{1}{180} \frac{n^3}{(1-n^2)} d_m^2\,.
\end{equation}

Although clay is highly porous, the presence of an electrical double-layer reduces its effective porosity resulting in reduced permeability. We use the effective porosity, considering the effect of the electrical double-layer, only for the permeability calculation, while the soil density is calculated based on the actual porosity and assuming the pore space is fully saturated with water. The saturated soil density is determined as:
\begin{equation}
 \rho_{sat} = \frac{(G_s + e) \rho_w}{1+e} \text{ and } n = \frac{e}{1+e}\,,   
\end{equation}
where $G_s$ is the specific gravity of solids, $e$ is the void ratio, $\rho_w$ is the density of water \SI{1000}{\kilogram\per\meter\cubed}, and $n$ is the soil porosity.~\Cref{tab:soil} presents the hydraulic and mechanical parameters of the soils used in this study. 

\begin{table}[!tb]
\caption{Soil material properties}
\label{tab:soil}
\begin{tabular}{llcccc}
\toprule
\textbf{Properties}           &   \textbf{units}                      & \textbf{Gravel} & \textbf{Sand} & \textbf{Silt} & \textbf{Clay} \\
\midrule
Porosity ($n$)                  & -                       & 0.3 - 0.35      & 0.3 - 0.45    & 0.4 - 0.45    & 0.45 - 0.6    \\
Grain size ($d\_m$)             & \si{\milli\meter}                      & 7.5 - 20        & 0.1 - 2.5     & 0.025-0.07    & 0.002 - 0.02  \\
Specific gravity ($G_s$)         & -                       & 2.75            & 2.65          & 2.67          & 2.7           \\
Permeability ($k$)              & \si{\meter\squared}    & 1E-7 - 1E-8     & 1E-9 - 1E-11  & 1E-12 - 1E-13 & 1E-13 - 1E-16 \\
Density ($\rho$) & \si{\kilo\gram\per\meter\cubed} & 2100 - 2200     & 1870 - 2110   & 1880 - 1970   & 1650 - 1890  \\
\bottomrule
\end{tabular}
\end{table}

\Cref{tab:mat} presents the thermal properties of the fluid, soil, and cables. We simulate the heat transfer of pipeline carrying water. The thermal conductivity of soil depends on its water content~\cite{young2001utilizing}. The bulk thermal conductivity and bulk specific heat of the porous media are calculated based on the volume fraction of fluid and solids:
\begin{align}
\lambda_b &= \lambda_s \cdot (1 - n) + \lambda_f  \cdot  n\,,\\
C_{p_b} &= C_{p_s}  \cdot (1 - n) + C_{p_f}  \cdot  n\,.
\end{align}

The thermal diffusivity of the soil is:
\begin{equation}
\alpha_s = \lambda_b / (\rho  \cdot  C_{p_b})\,.
\end{equation}

\begin{table}[htbp]
\caption{Thermal properties of the fluid, soil, and the cables}
\label{tab:mat}
\centering
\begin{tabular}{ll}
\toprule
\textbf{Properties}                         & \textbf{Value}   \\
\midrule
\textbf{Fluid}                              &                  \\
Fluid density ($\rho_f$)                    & 1000 \si{\kilo\gram\per\meter\cubed}    \\
Specific heat of fluid ($Cp_f$)             & 4290 \si{\joule\per\kilogram-\kelvin}    \\
Thermal conductivity of fluid ($\lambda_f$) & 0.6 \si{\watt\per\meter-\kelvin} \\
Thermal diffusivity of fluid ($\alpha_f$)   & 1.39E-7 \si{\meter\squared\per\second}  \\
Dynamic viscosity of fluid ($\mu$)          & 1E-3 \si{\pascal\cdot\second}      \\
\textbf{Soil}                               &                  \\
Specific heat of soil ($C_{p_s}$)           & 8000 \si{\joule\per \kilogram-\kelvin}    \\
Thermal conductivity of soil ($\lambda_s$)  & 1.0 \si{\watt\per\meter-\kelvin}      \\
\textbf{Pipe}                               &                  \\
Thermal diffusivity of pipe ($\alpha_p$)    & 1.172E-5 \si{\meter\squared\per\second}  \\
\bottomrule
\end{tabular}
\end{table}

\pagebreak
\section{Results and discussion}
We investigate the heat transfer from the HV piggyback cables to the pipeline and surrounding soil for different site conditions. We bury the HV cables to a depth of \SI{0.5}{\meter} from the ground surface and the three cables are heated to an above ambient temperature of \SI{70}{\celsius}. We first discuss the distribution of heat around the cables and the pipeline in soil with a permeability $k$ of \SI{1.67E-11}{\meter\squared}.  Subsequently, we investigate how permeability influences the heat distribution identifying a critical Rayleigh number that delineates conduction from convection-dominated flows. 

\subsection*{Time evolution of heat distribution}
As the cables are heated, the temperature in the pipeline and surrounding soil gradually rises, eventually reaching a steady state which shows no further change in temperature.~\Cref{fig:time-evol} shows the time evolution of temperature distribution around a pipe heated by three HV piggy-backed cables buried to a depth of~\SI{0.5}{\meter} in sand with a permeability $k$ of \SI{1.67E-11}{\meter\squared} ($Ra = 7.29$). In the first 10 hours, the cables heat the pipe, resulting in a linear temperature gradient in the pipe wall from the crown to the pipe invert. Whereas, the temperature of the fluid in the pipe increases non-uniformly, with the fluid at the center having the lowest temperature. Between 10 and 100 hours, the fluid inside the pipe begins to heat more uniformly throughout the cross-section, achieving 95\% of its steady-state temperature. The temperature increase decelerates between 100 and 200 hours, as the pipe temperature reaches 99.2\% of the steady-state condition. The pipe temperature reaches the steady-state beyond 700 hours, while the temperature of the pipe fluid reaches steady-state beyond 900 hours. 

\begin{figure}
\centering
  \begin{subfigure}[b]{0.49\textwidth}
    \includegraphics[width=\textwidth]{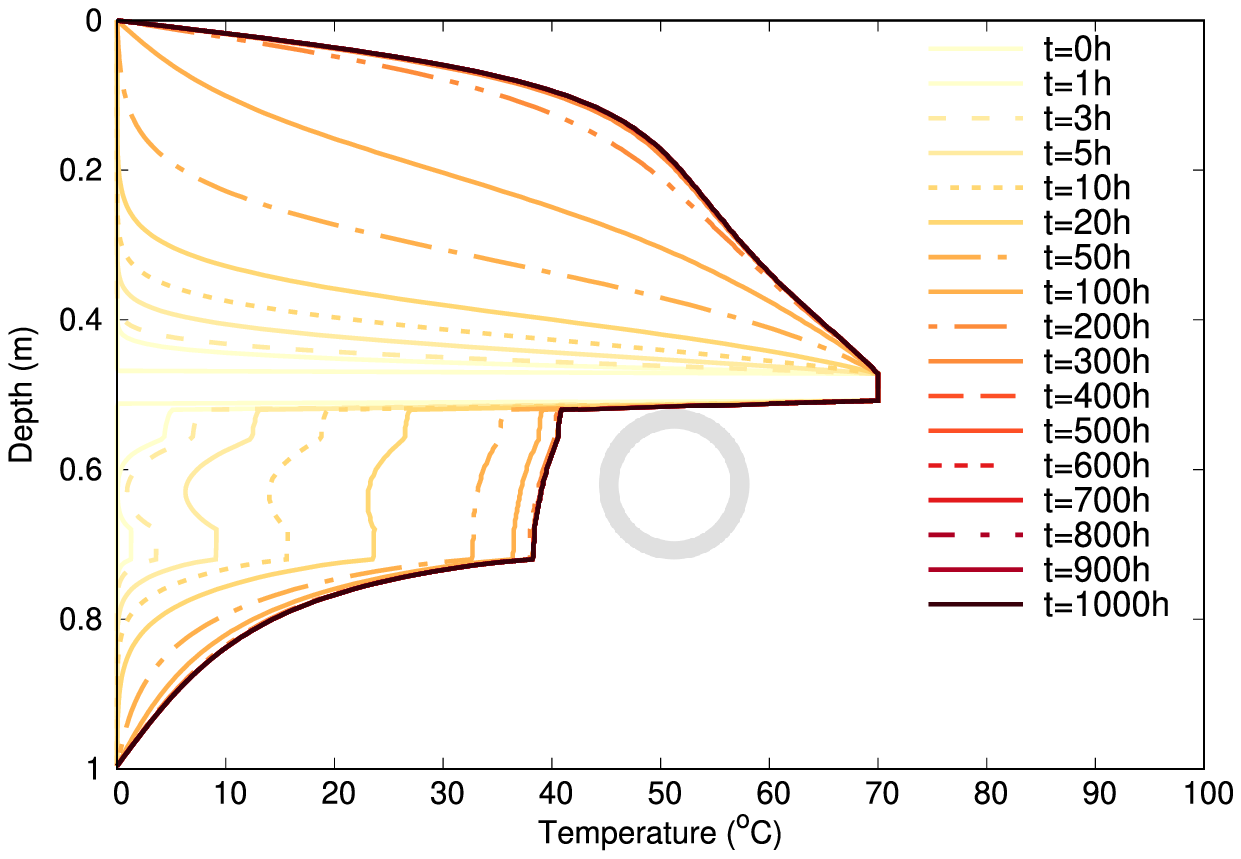}
    \caption{the vertical axis (A-A’)}
    \label{fig:temp-vline}
 \end{subfigure}             
 \begin{subfigure}[b]{0.49\textwidth}
    \includegraphics[width=\textwidth]{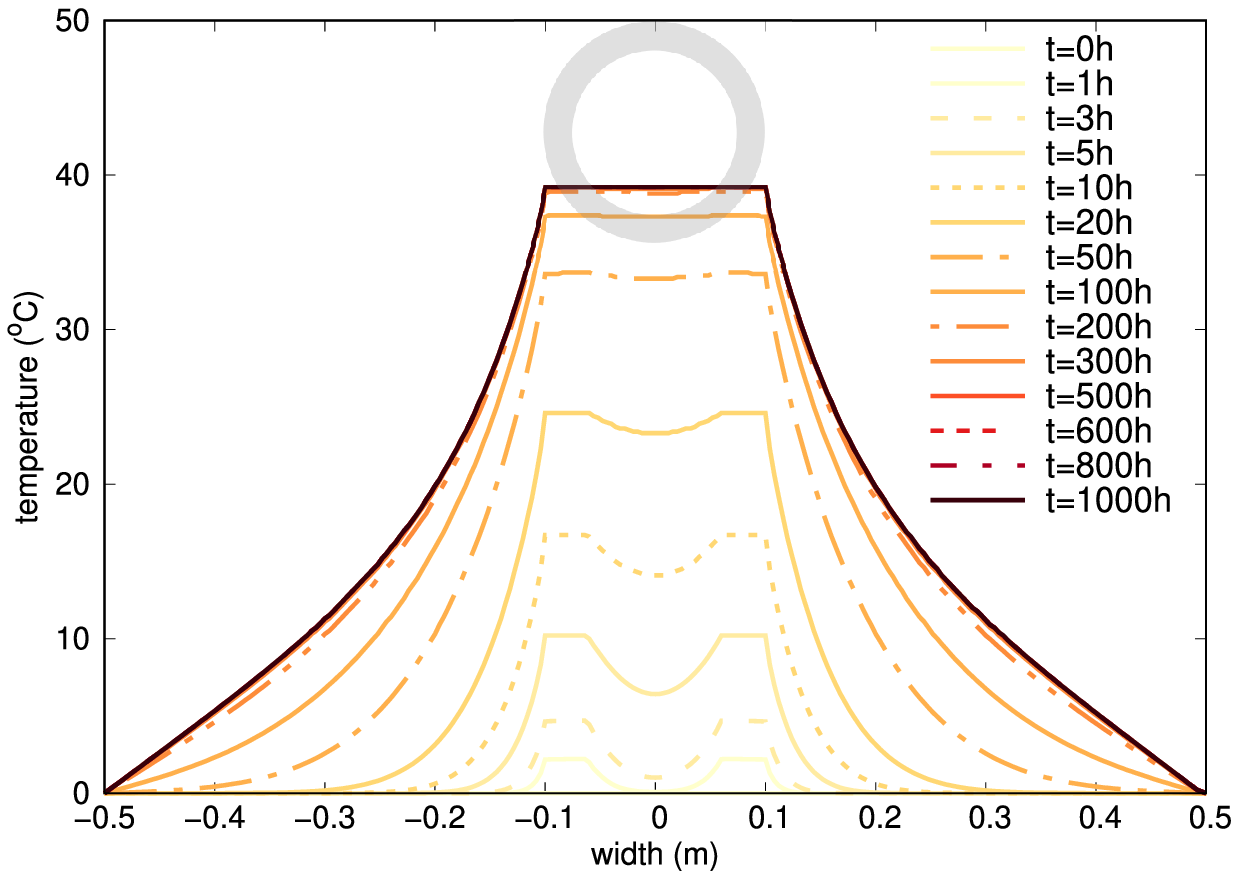}
    \caption{the horizontal axis (B-B’)}
    \label{fig:temp-hline}
 \end{subfigure}
 
 \caption{Evolution of temperature with time along the vertical (A-A’) and horizontal (B-B’) axis along the pipe center for a soil with $k = \SI{1.67E-11}{\meter\squared} (Ra = 7.29)$. The cables are heated to an above ambient temperature of \SI{70}{\celsius} and the burial depth is \SI{0.5}{\meter}.}
 \label{fig:temp-depth-hline}
\end{figure}

Conductive heat transfer creates a symmetric spatial distribution of heat around the cable, while convection breaks the symmetry with a preferential upward flow. As the temperature distribution evolves with time (see \cref{fig:time-evol}), the heat transfer around the pipe remains symmetric for 20 hours, characteristic of a conductive behavior. After 50 hours, convection starts to dominate as buoyant forces drive the heat towards the ground surface resulting in asymmetric heat distribution in the vertical direction.~\Cref{fig:temp-vline} shows the temperature distribution with depth along the vertical axis of the pipe (AA’ in~\cref{fig:model}) at different times. For the entire duration, the amount of temperature increase below the pipe (0.72 to \SI{1}{\meter} from the ground surface) is lower than the temperature increase above the pipe (0 to \SI{0.45}{\meter}). After 200 hours, the average temperature above the pipe is \SI{46.4}{\celsius}, more than four times the heat below the pipe (\SI{10.8}{\celsius}). The asymmetric distribution of heat above and below the pipe is indicative of a convection-dominated flow. Convection carries heat away from the pipe towards the ground surface above the cables, while the heat dissipation below the pipe is predominantly through conduction. The heat distribution below the pipe is similar to the symmetric distribution seen in \cref{fig:temp-hline} along the horizontal direction, where conduction predominates.

\begin{figure}
 \centering
 \begin{subfigure}[b]{0.49\textwidth}
    \includegraphics[width=\textwidth]{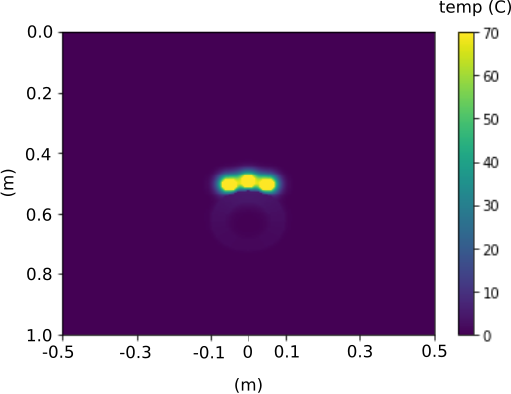}
    \caption{\SI{1}{\hour}}
 \end{subfigure}             
 \begin{subfigure}[b]{0.49\textwidth}
    \includegraphics[width=\textwidth]{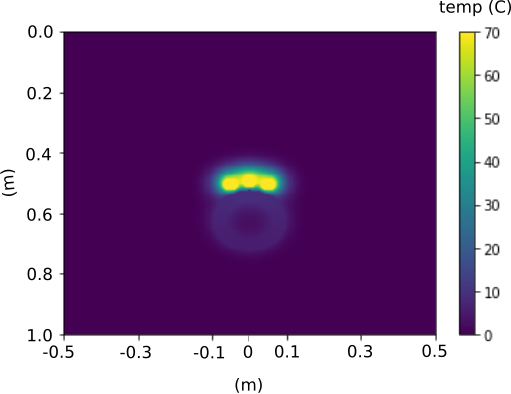}
    \caption{\SI{3}{\hour}}
 \end{subfigure}\\           
 \begin{subfigure}[b]{0.49\textwidth}
    \includegraphics[width=\textwidth]{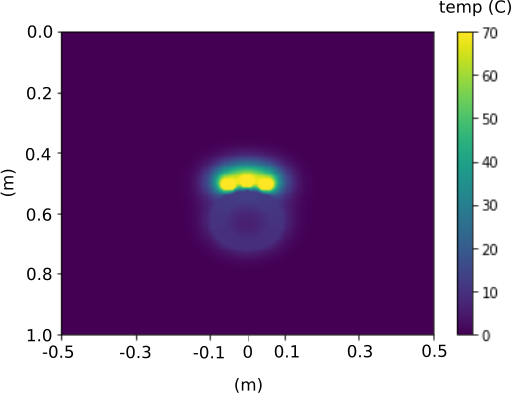}
    \caption{\SI{5}{\hour}}
 \end{subfigure}             
 \begin{subfigure}[b]{0.49\textwidth}
    \includegraphics[width=\textwidth]{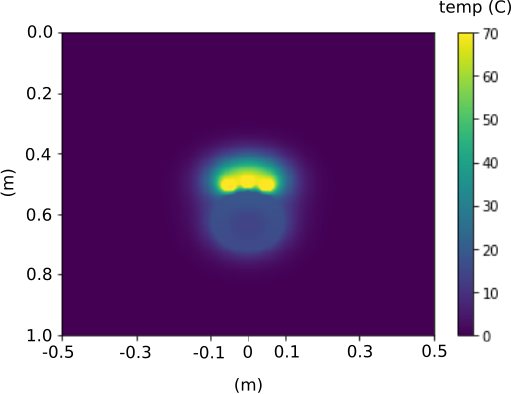}
    \caption{\SI{10}{\hour}}
 \end{subfigure}   \\
 \begin{subfigure}[b]{0.49\textwidth}
    \includegraphics[width=\textwidth]{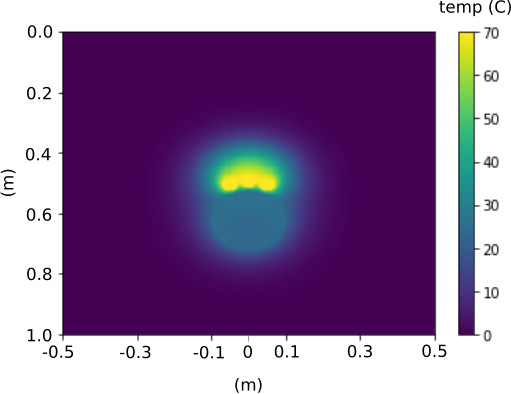}
    \caption{\SI{20}{\hour}}
 \end{subfigure}             
 \begin{subfigure}[b]{0.49\textwidth}
    \includegraphics[width=\textwidth]{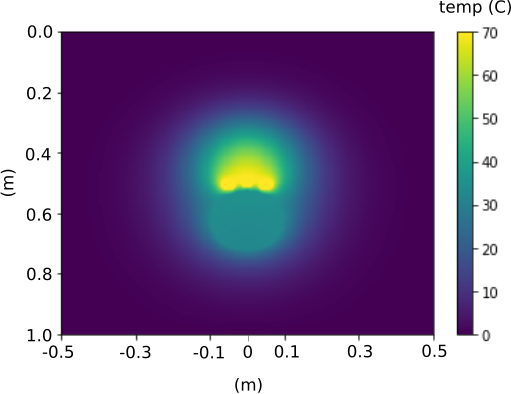}
    \caption{\SI{50}{\hour}}
 \end{subfigure}\\           
\end{figure}%
\begin{figure}[ht]\ContinuedFloat 
 \begin{subfigure}[b]{0.49\textwidth}
    \includegraphics[width=\textwidth]{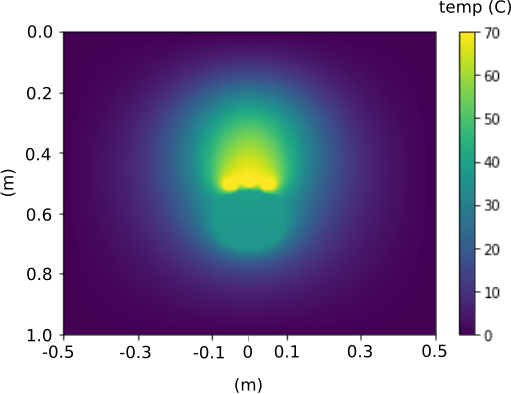}
    \caption{\SI{100}{\hour}}
 \end{subfigure}             
 \begin{subfigure}[b]{0.49\textwidth}
    \includegraphics[width=\textwidth]{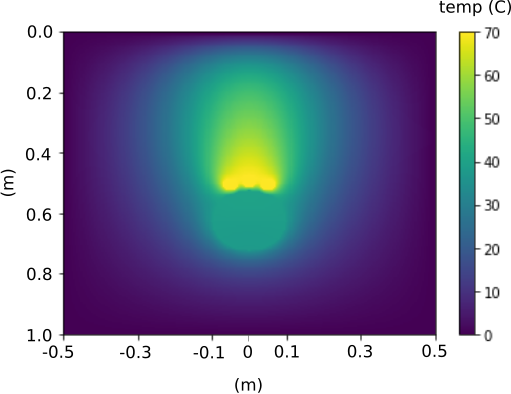}
    \caption{\SI{200}{\hour}}
    \label{fig:200}
 \end{subfigure}  \\
  \begin{subfigure}[b]{0.49\textwidth}
    \includegraphics[width=\textwidth]{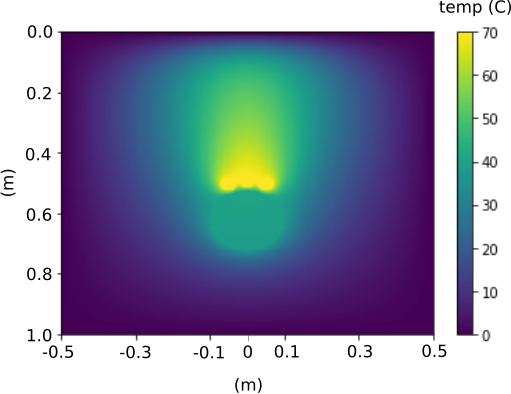}
    \caption{\SI{500}{\hour}}
    \label{fig:500}
 \end{subfigure}             
 \begin{subfigure}[b]{0.49\textwidth}
    \includegraphics[width=\textwidth]{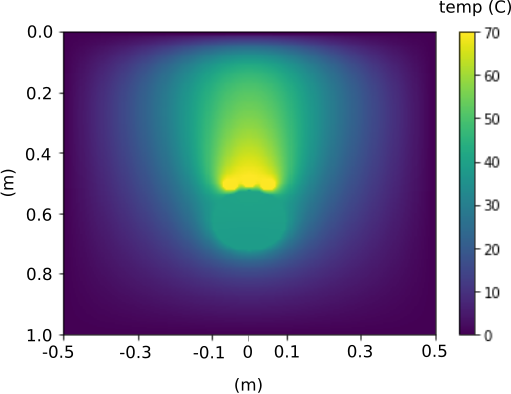}
    \caption{\SI{1000}{\hour}}
    \label{fig:1000}
 \end{subfigure}  \\
 
 \caption{Evolution of temperature with time in the region of interest (as shown in~\cref{fig:model}) for soil with a permeability $k = \SI{1.67E-11}{\meter\squared} (Ra = 7.29)$. The cables are heated to \SI{70}{\celsius} above ambient temperature and the burial depth is \SI{0.5}{\meter}.}
 \label{fig:time-evol}
\end{figure}

\begin{figure}
    \centering
    \includegraphics{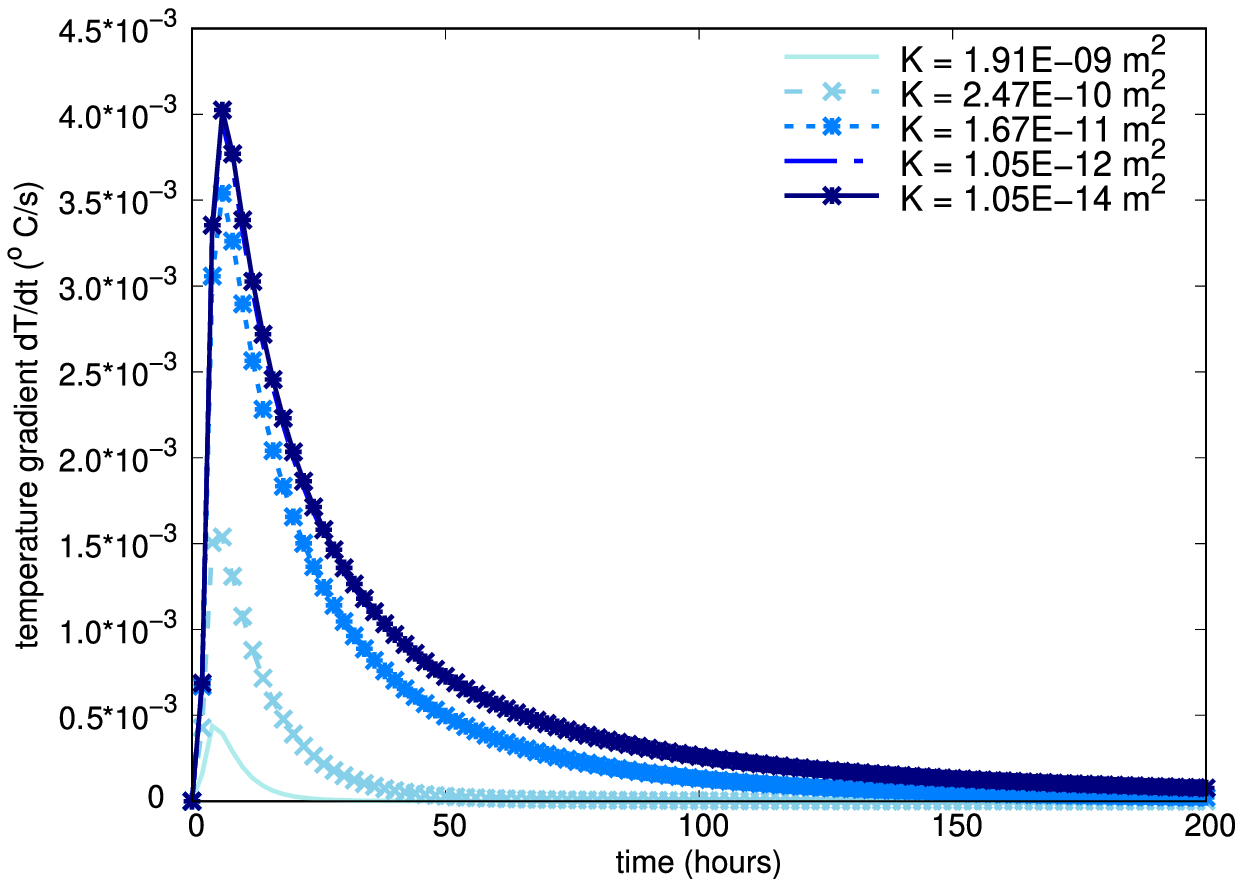}
    \caption{Evolution of temperature gradient in different soil permeabilities. The cables are heated to an above ambient temperature of \SI{70}{\celsius} and the burial depth is \SI{0.5}{\meter}.}
    \label{fig:dTdt}
\end{figure}

\subsection*{Permeability variation}
The Rayleigh-Darcy dimensionless number ($Ra$) represents the ratio of time scales between conductive and convective flow in the soil. The critical Rayleigh-Darcy number, $Ra_c$, that differentiates regions dominated by conduction and convection varies based on the boundary conditions. Unlike other factors defining the Rayleigh-Darcy number, see~\cref{eq:ra}, the permeability of natural soils ($k$) varies by several orders of magnitude from \SI{1E-16}{\meter\squared} in clayey soil to \SI{1E-7}{\meter\squared} in sandy materials. Hence, of the various factors influencing the Rayleigh-Darcy number, the soil permeability is a critical factor in determining the dominant mode of heat transfer -- conduction or convection.

We vary the soil permeability from \SI{1E-7} to \SI{1E-16}{\meter\squared} to determine the effect of permeability on the mode of heat dissipation.~\Cref{fig:dTdt} shows the rate of temperature change with time at the center of the pipe buried in soils with different permeabilities. For high permeability soils ($k > \SI{1E-10}{\meter\squared}$), the steady-state condition is reached within 50 hours, whereas low permeability soils ($k < \SI{1E-12}{\meter\squared}$) show no noticeable change in temperature gradients beyond 150 hours.~\Cref{fig:200,fig:500,fig:1000} shows the heat distribution remains unchanged beyond 200 hours. We choose 200 hours for all subsequent simulations to balance the long computational duration and achieve a system closer to steady-state. 

\Cref{fig:ra-evol} shows the temperature distribution around the pipe after 200 hours in different soil permeabilities. At high permeabilities ($k > \SI{2E-10}{\meter\squared}$), convection dominates, resulting in a near-vertical transfer of heat away from the cables and barely heating the pipe. The seepage velocity in soil controls the time-scale of convection. ~\Cref{fig:dTdt} shows high permeability soils reaching the steady-state temperature sooner ($t < 25$ hours) than low permeability soils, where steady-state is reached at $t > 700$ hours. In high permeability soils, the large-interconnected pore space results in higher seepage velocity and a lower time-scale of convection, i.e., higher Rayleigh number. Buoyancy forces drive fluid circulation in high permeability soils as the ease of fluid flow improves, reducing the time-scale for convection. Convection-driven flows have higher Rayleigh numbers and a preferential upward migration of heat away from the source. Convective flow is detrimental to inductive heating of the pipeline due to significant loss of heat to the saturated soil media above the cables.

As the permeability decreases from $k =$ \SI{4.9E-10} to \SI{1.6E-11}{\meter\squared}, the heat distributes more uniformly around the cable, heating the pipe and the fluid; but still showing a preferential upward heat distribution reaching the ground surface typical in a convection-dominated behavior. The pipe temperature increases showing symmetric heat distribution around the cable as the permeability decreases from \SI{1E-12} to \SI{1E-15}{\meter\squared}. In these low permeability conditions, conduction dominates, and the pipeline is heated efficiently, with minimal heat loss to the surrounding soil. At very low permeabilities (in silty and clayey soil with $k < \SI{1E-12}{\meter\squared}$ and $d_m < \SI{70}{\micro\meter}$), the narrow and sparsely interconnected pore spaces limit the fluid flow velocity. Hence, the time scale of convection is much larger than that of conduction, i.e., a smaller Rayleigh number indicating conduction-driven heat transfer. Conductive heat transfer causes an approximately isotropic distribution of heat around the source. 

For a given burial depth, the density of soil determines the time scale of conduction. Unlike the enormous variation in the soil permeability with grain size, the soil density only varies in a narrow range. Although this variation in density affects the time scale of conduction, the difference in the temperature distribution for low permeability soils is negligible at steady state.

\begin{figure}
 \centering
 \begin{subfigure}[b]{0.49\textwidth}
    \includegraphics[width=\textwidth]{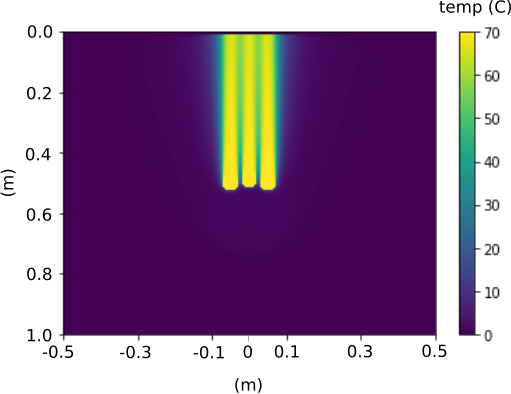}
    \caption{$Ra$ = 955, $k$ = \SI{1.91E-9}{\meter\squared}}
 \end{subfigure}
 \begin{subfigure}[b]{0.49\textwidth}
    \includegraphics[width=\textwidth]{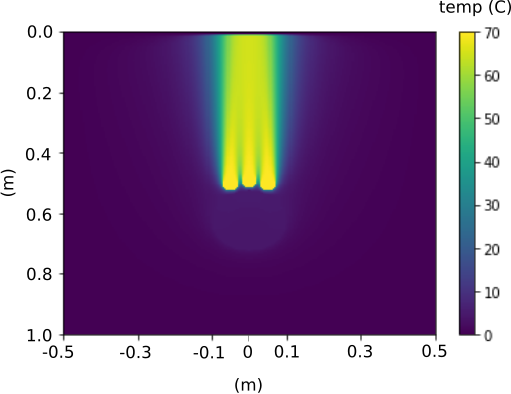}
    \caption{$Ra$ = 269.5, $k$ = \SI{5.64E-10}{\meter\squared}}
 \end{subfigure}\\
 \begin{subfigure}[b]{0.49\textwidth}
    \includegraphics[width=\textwidth]{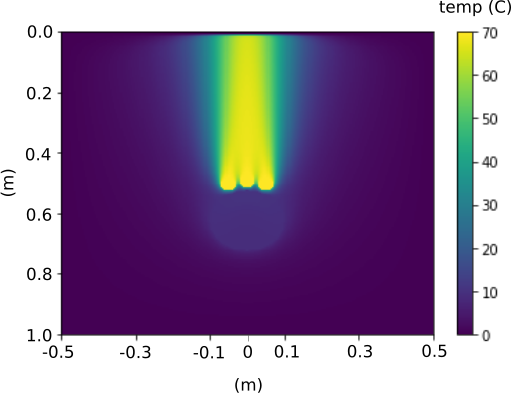}
    \caption{$Ra$ = 112, $k$ = \SI{2.47E-10}{\meter\squared}}
 \end{subfigure}
 \begin{subfigure}[b]{0.49\textwidth}
    \includegraphics[width=\textwidth]{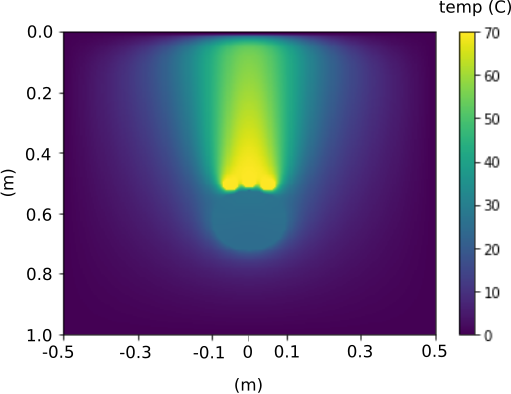}
    \caption{$Ra$ = 21.95, $k$ = \SI{4.89E-11}{\meter\squared}}
 \end{subfigure}\\ 
 \begin{subfigure}[b]{0.49\textwidth}
    \includegraphics[width=\textwidth]{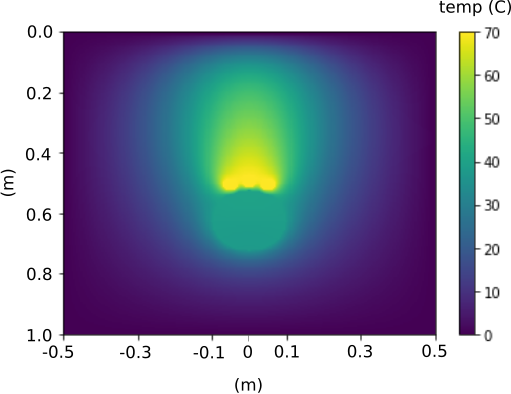}
    \caption{$Ra$ = 7.29, $k$ = \SI{1.67E-11}{\meter\squared}}
 \end{subfigure}
 \begin{subfigure}[b]{0.49\textwidth}
    \includegraphics[width=\textwidth]{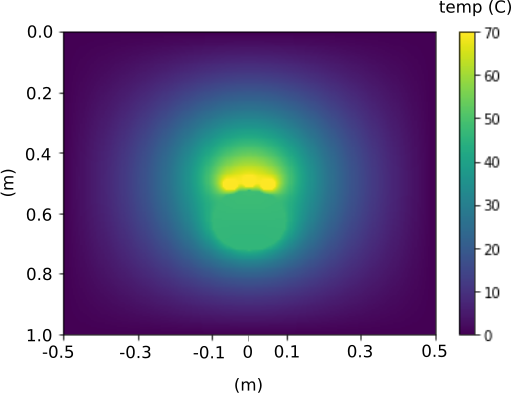}
    \caption{$Ra$ = 2.225, $k$ = \SI{4.84E-12}{\meter\squared}}
 \end{subfigure}\\  
\end{figure}%
\begin{figure}[ht]\ContinuedFloat 
 \begin{subfigure}[b]{0.49\textwidth}
    \includegraphics[width=\textwidth]{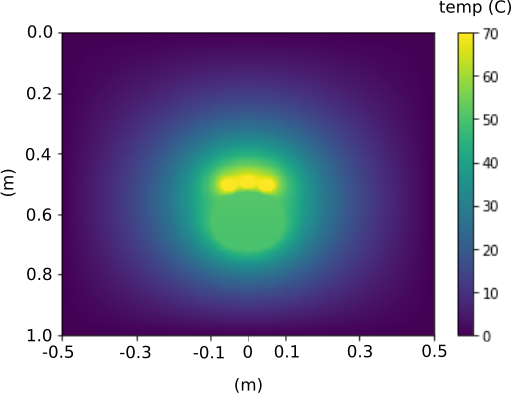}
    \caption{$Ra$ = \SI{4.58E-1}, $k$ = \SI{1.05E-12}{\meter\squared}}
 \end{subfigure}
 \begin{subfigure}[b]{0.49\textwidth}
    \includegraphics[width=\textwidth]{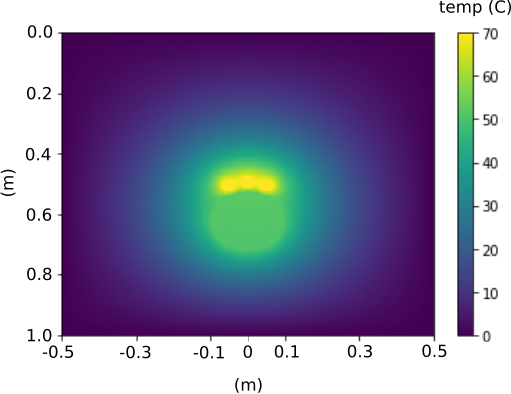}
    \caption{$Ra$ = \SI{4.9E-2}, $k$ = \SI{1.11E-13}{\meter\squared}}
 \end{subfigure}\\    
 \begin{subfigure}[b]{0.49\textwidth}
    \includegraphics[width=\textwidth]{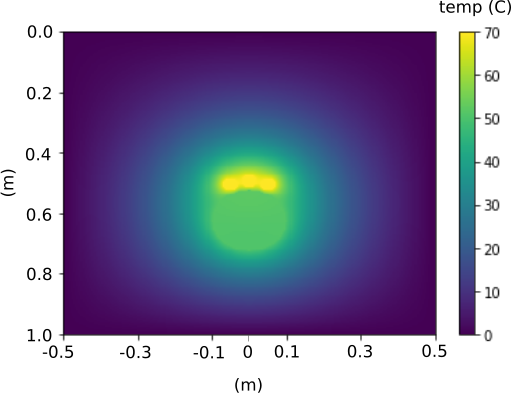}
    \caption{$Ra$ = \SI{4.4E-4}, $k$ = \SI{1.05E-14}{\meter\squared}}
 \end{subfigure}
 \begin{subfigure}[b]{0.49\textwidth}
    \includegraphics[width=\textwidth]{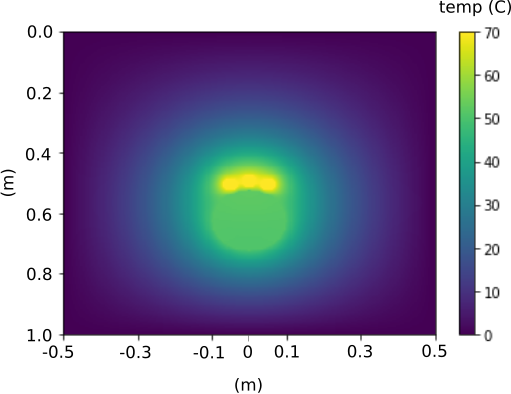}
    \caption{$Ra$ = \SI{4.4E-4}, $k$ = \SI{1.18E-15}{\meter\squared}}
 \end{subfigure}\\ 
 \caption{Temperature distribution in different soil permeability conditions after 200 hours. The cables are heated to an above ambient temperature of \SI{70}{\celsius} and the burial depth is \SI{0.5}{\meter}.}
 \label{fig:ra-evol}
\end{figure}

The heat transfer mechanism changes from conduction to convection-dominated as the permeability of soil increases. A critical Rayleigh-Darcy number ($Ra_c$) defines this transition point from conduction to convection-dominated region.~\citeN{nield1968onset} defines this transition at a critical $Ra$ of 12 for a rectangular domain with a constant heat source with Free-Free boundary condition. At low permeabilities, where $Ra < 1$, the temperature field is symmetric, indicating a conductive behavior (see~\cref{fig:ra-evol} g-j). The asymmetric convective behavior occurs for $Ra > 2.22$ ($k > \SI{4.84E-12}{\meter\squared}$). The observation of convective response for $Ra = 2.22$, below the critical $Ra$ of 12, indicates the transition occurs at lower $Ra_c$ than originally thought. 

\begin{figure}
    \centering
    \includegraphics{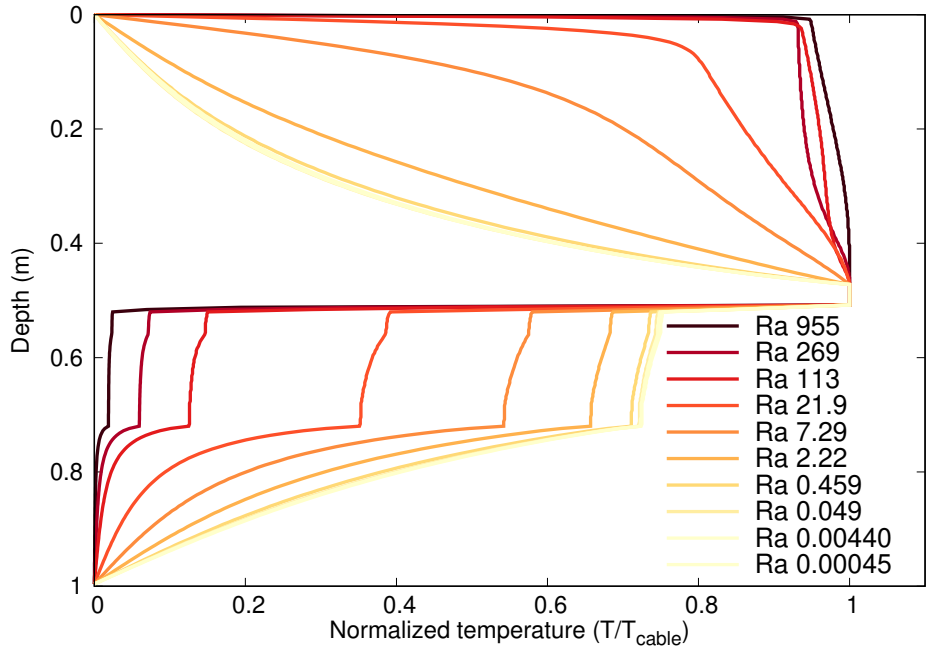}
    \caption{Temperature distribution along the vertical axis (A-A’) through the center of pipe for different Rayleigh-Darcy numbers.}
    \label{fig:ra-vline}
\end{figure}

\Cref{fig:ra-vline} shows the temperature distribution with depth along the A-A' axis for different Rayleigh-Darcy numbers. The fluid temperature in the pipe reaches a maximum of 73.5\% of the cable temperature ($T_{cable}$) for the lowest $Ra = $~\SI{4.5E-4} ($k =$ \SI{1.18E-15}{\meter\squared}). The fluid temperature inside the pipe remains almost unchanged with increasing Rayleigh-Darcy number until a $Ra = 0.46$ ($k >$ \SI{1E-12}{\meter\squared}), above which the fluid temperature drops drastically. For large Rayleigh-Darcy numbers ($Ra > 100$), the fluid in the pipe barely reaches 5\% of the cable temperature. Below the pipe, the heat dissipation profile remains unchanged with increasing $Ra$, which indicates conductive behavior dominating the region below the pipe. The heat distribution above the cable varies as the Rayleigh-Darcy number increases due to buoyancy forces. Lower $Ra$ conditions show a lower rate of dissipation (concave profile), whereas soils experiencing a higher Rayleigh-Darcy number show an accelerated heat distribution (convex profile). The heat profile above the cables transitions from concave to convex between a Rayleigh-Darcy number of 2 and 7, below the established critical Rayleigh-Darcy ($Ra_c$) of 12~\cite{nield1968onset}. 

\begin{figure}
    \centering
    \includegraphics{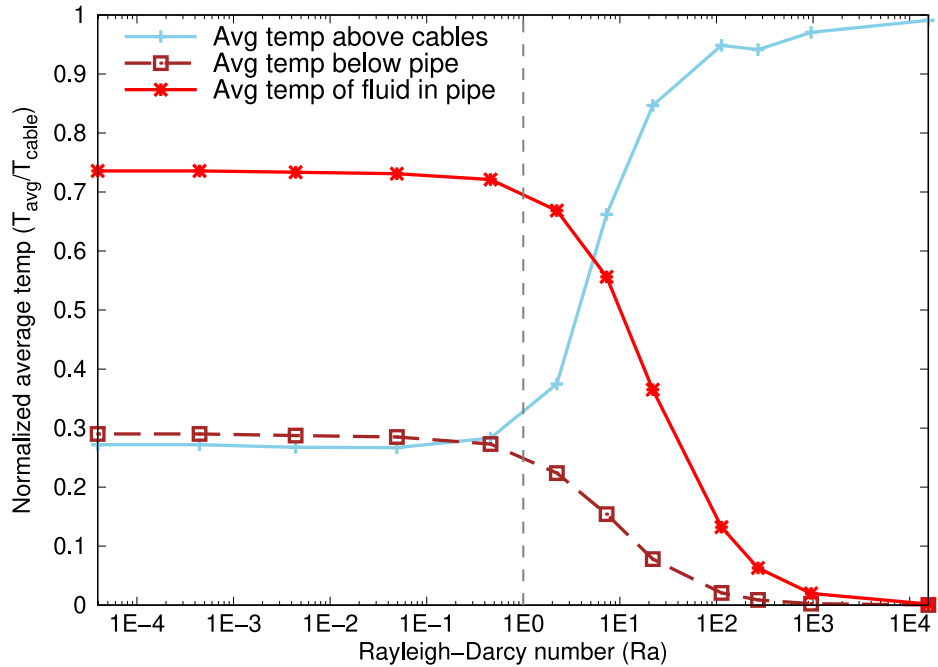}
    \caption{The average temperature of fluid in pipes buried in soils with different Rayleigh-Darcy numbers.}
    \label{fig:heat-pipe}
\end{figure}

We closely examine the heat distribution above and below the pipeline to determine different modes of heat transfer.~\Cref{fig:heat-pipe} shows the evolution of normalized average temperatures of the fluid in the pipe, the soil below the pipe, and soil above the cables for different Rayleigh-Darcy numbers. For $Ra < 1$, the average pipe fluid temperature remains unchanged at 73.5\% of the cable temperature, and the soil temperature above the cables stays constant at 28\% of the cable temperature. As the Rayleigh-Darcy number increases from 1 to 100, the temperature of the pipe fluid drastically drops to 13\%, while the soil temperature above the cables rises to 95\% of cable temperature. The heating of pipe using cables is nonexistent for $Ra > 100$, where convection transfers almost all of the cable heat to the soil above. Hence, the transition from conduction to convection happens below a Rayleigh-Darcy number of 1.0.

\begin{figure}
    \centering
    \includegraphics{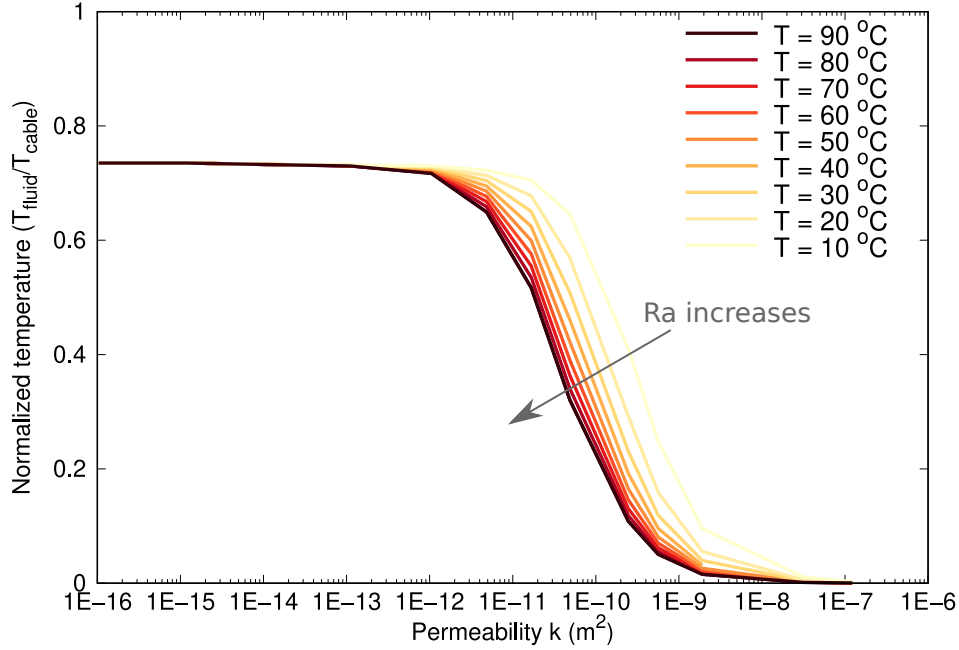}
    \caption{Normalized pipe temperature for different soil permeabilities.}
    \label{fig:k-temp}
\end{figure}

In addition to soil permeability, the temperature in the cable also influences the Rayleigh-Darcy number, and in turn, the transition from conduction to convection-dominated heat transfer mechanism. We vary the temperature in the cable from \SI{5}{\celsius} to \SI{90}{\celsius} for different soil permeabilities.~\Cref{fig:k-temp} summarizes the combined influence of temperature and permeability on the average fluid temperature in the pipe. We use the normalized pipe fluid temperature ($T_{fluid}/T_{cable}$) to investigate how the cable temperature influences the heat transfer mechanism at different permeabilities. 

For low permeabilities ($k < \SI{1E-12}{\meter\squared}$), irrespective of the applied cable temperature, the pipe temperature plateaus at a constant maximum temperature of 73.5\% of the cable temperature. The constant pipe temperature in low permeability soil indicates that conduction is the primary heat transfer mechanism. A transition region exists for a permeability range of \SI{1E-12} to \SI{1E-9}{\meter\squared}, where conduction and convection play a role in distributing the heat from the cables. In the transition zone, the efficiency of heating decreases as the permeability increases—the proportion of convection to conduction controls the heat transfer efficiency. For a given permeability, the lower the temperature higher is the heat transfer efficiency. In other words, as the cable temperature increases, more heat is lost to the ground surface due to convection, and the proportion of heat transferred to the pipe drops. An increase in cable temperature increases the Rayleigh-Darcy number proportionally, see~\cref{eq:ra}, thus favoring convection over conduction. As the permeability increases beyond \SI{1E-9}{\meter\squared}, convection dominates, driving the heat away from the cable towards the ground surface, thus barely heating the pipe irrespective of cable temperature. Hence, the Rayleigh-Darcy number, and not permeability, is the controlling factor in determining the mode of heat transfer.

\begin{figure}
    \centering
    \includegraphics{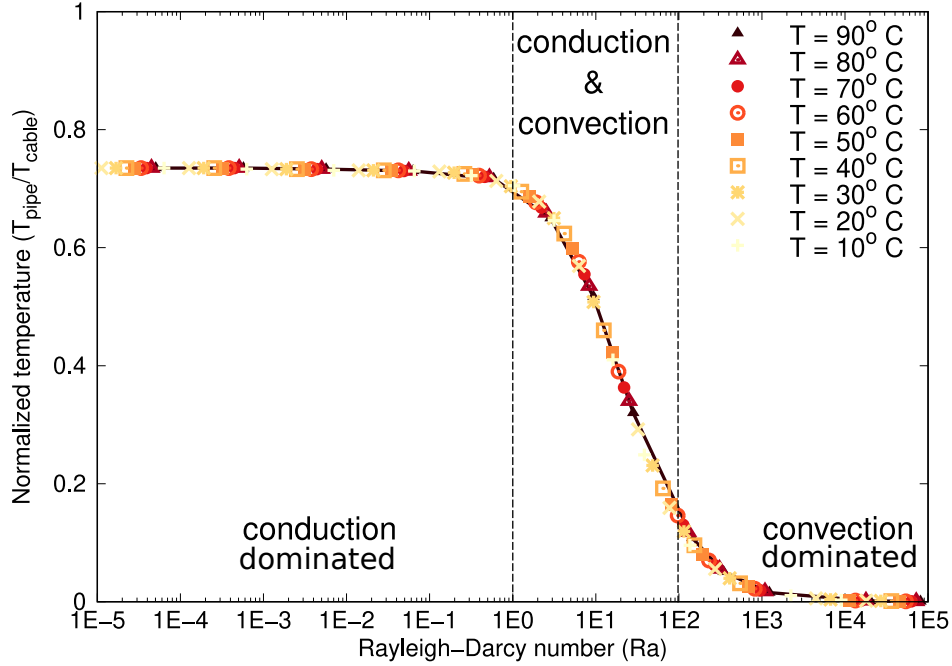}
    \caption{Normalized pipe temperature for different Rayleigh-Darcy numbers.}
    \label{fig:ra-temp}
\end{figure}

We replot the normalized cable temperature for different soil permeabilities as a function of the Rayleigh-Darcy number (see~\cref{fig:ra-temp}) to show the dependence of heat transfer on the Rayleigh-Darcy number. The normalized pipe temperature across all permeabilities collapses onto a single curve. The normalized pipe temperature remains constant when conduction dominates ($Ra < 1$). In the conduction region, the heat transfer has a maximum efficiency of 73.5\% for a burial depth of \SI{0.5}{\meter}. As the Rayleigh-Darcy number increases between 1 and 100, the efficiency of heat transfer drops drastically. An increase in the Rayleigh-Darcy number means an increase in convection, driving heat away from the cables towards the ground surface. Convection is the dominant heat transfer mechanism at higher Rayleigh-Darcy numbers ($Ra > 100$) and has the lowest heat transfer efficiency. 

\begin{figure}
    \centering
    \includegraphics[width=\textwidth]{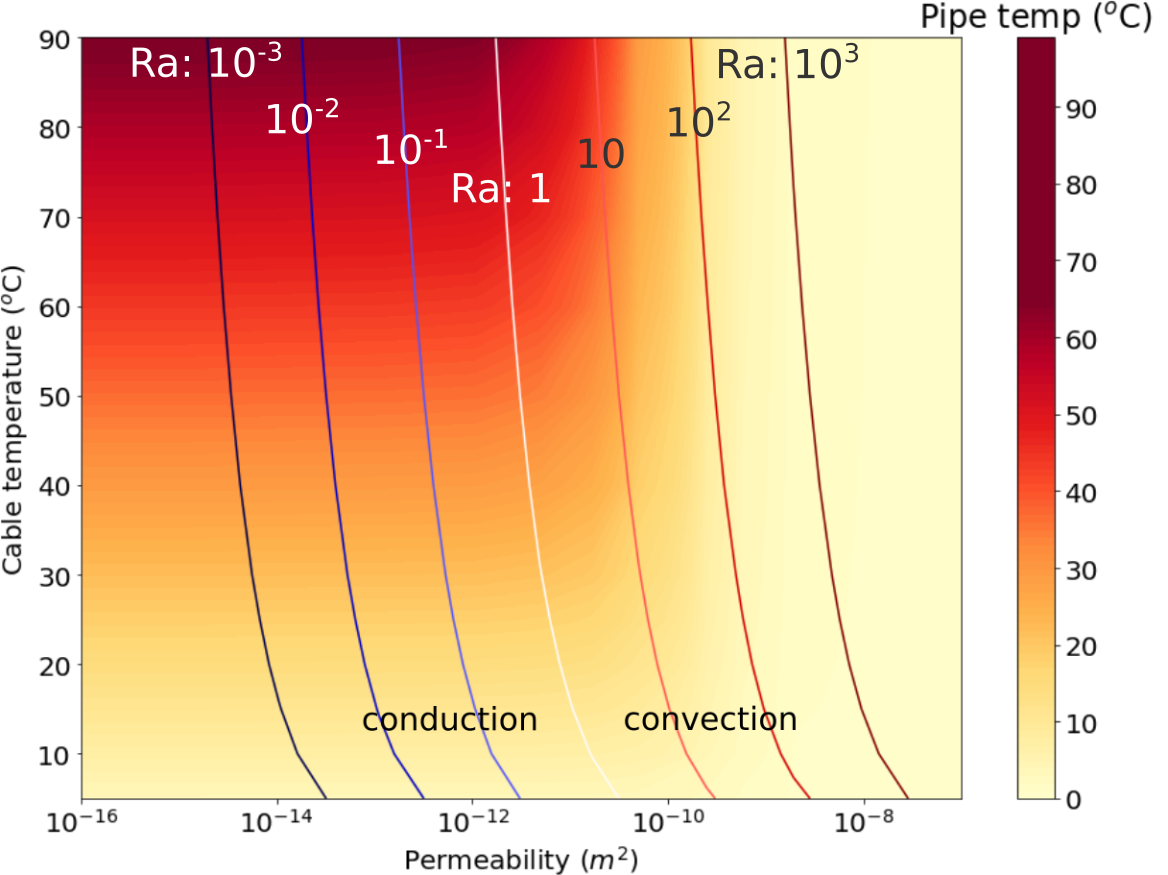}
    \caption{Variation of the pipe fluid temperature for different soil permeabilities and cable temperatures. The solid curves show different Rayleigh-Darcy numbers increase from left to right with an increase in permeability.}
    \label{fig:ra-temp-contour}
\end{figure}

\Cref{fig:ra-temp-contour} shows the variation in the average pipe fluid temperature for different soil permeabilities and cable temperatures. Conduction predominates for low permeability soils ($k < \SI{1E-12}{\meter\squared}$). With negligible convection, there is no heat loss away from the cables. Conversely, for very high permeability soils ($k > \SI{1E-10}{\meter\squared}$), majority of the heat is transferred away from the cables by convection. The onset of convective behavior also depends on the cable temperature. Convection increases with an increase in the cable temperature. We identify a transition at a $Ra_c$ of 1, which separates regions dominated by conduction ($Ra_c < 1$) and convection ($Ra_c > 1$). Above an $Ra$ of 100, only convective flow occurs, and the pipe remains unheated. 

\subsection*{Relationship between heat transfer and flow condition}
The Rayleigh-Darcy number ($Ra$) defines the amount of free convective flow, but not the amount of heat transfer. Whereas, the Nusselt number ($Nu$) defines the ratio of convective and conductive heat transfer at a boundary. In regions of natural convective heat transfer, a power law relates the Rayleigh-Darcy number ($Ra$) to the Nusselt number for convection ($Nu_{conv}$) ~\cite{hardee1976boundary,merkin1979free}:
\begin{equation}
Nu_{conv} = 0.565 * Ra^{0.5}\,.
\label{eq:nu-conv}
\end{equation}
The Nusselt number for conduction ($Nu_{cond}$) is a function of the burial depth:
\begin{equation}
    Nu_{cond} = \frac{2}{\ln \left(\frac{4 D}{d}\right)}\,,
    \label{eq:nu-cond}
\end{equation}
where $D$ is the depth of burial and $d$ is the diameter of the cable. 

The Nusselt number is defined as: 
\begin{equation}
Nu = \frac{hD}{\lambda}  = \frac{qD}{\Delta T \lambda}\,,
\end{equation}
where $h$ is the convective heat transfer coefficient (\si{\watt\per\meter\squared\kelvin}), $D$ is the characteristic length (burial depth), $\lambda$ is the thermal conductivity (\si{\watt\per\meter-\kelvin}), $q$ is the heat flux from the cable (\si{\watt}), and $\Delta T$ is the temperature difference between the cable surface and the ambient temperature. In the simulations, the Nusselt number is calculated as: 
\begin{equation}
Nu = \frac{hD}{\lambda} =  \frac{\frac{\partial (T_c - T)}{\partial y}|_{y=0}}{T_s -T_\infty}\,,
\end{equation}
where $T_c$ is the cable temperature and $T_\infty$ is the temperature near the ground surface. 

\begin{figure}
    \centering
    \includegraphics{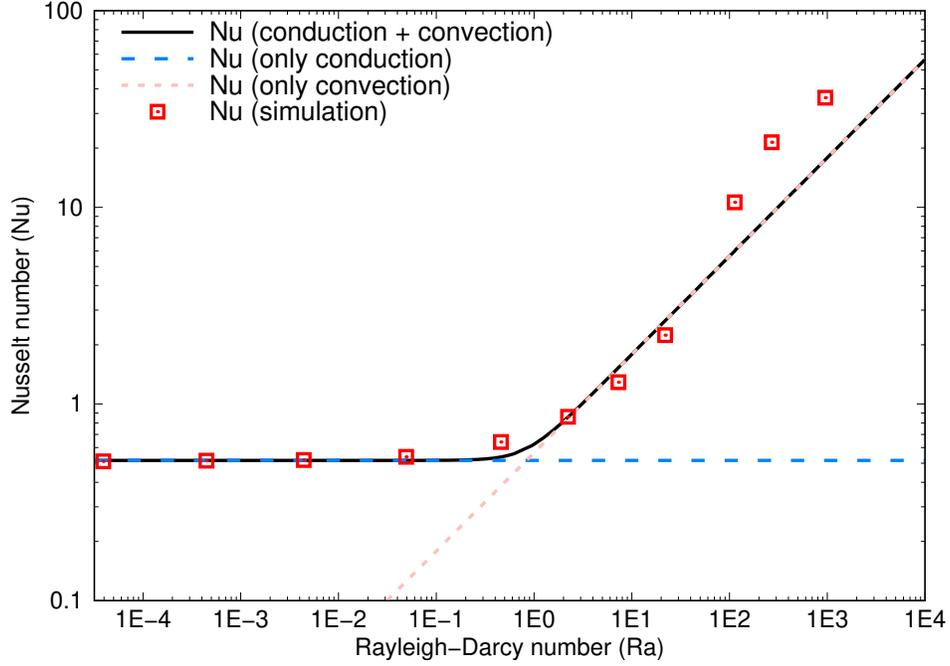}
    \caption{Variation of Nusselt number as a function of Rayleigh-Darcy number for only conduction, only convection and heat transfer with conduction and convection.}
    \label{fig:ra-nu}
\end{figure}

\Cref{fig:ra-nu} shows the relationship between the Nusselt number and the Rayleigh-Darcy number for different temperatures. The Nusselt number based on the simulation matches the Nusselt number for conduction (\cref{eq:nu-cond}) for low permeability conditions ($Ra < 1$). At high permeabilities ($Ra > 100$), the Nusselt number increases with the Rayleigh-Darcy number as a power of 0.5 as in the convection relation in~\cref{eq:nu-conv}. A transition region exists for Ra between 1 and 100, where a power-law relation describes the combined influence of conduction and convection. In the transition zone, the simulation Nusselt number ($Nu_{sim}$) shows that the effect of conduction gradually decreases with increasing Rayleigh-Darcy numbers, eventually reaching a fully convective flow and heat transfer.~\citeN{churchill1972general} expressed the heat transport process to vary uniformly between two limiting solutions of purely conductive and convective heat transfer:
\begin{equation}
    Nu = (Nu_{cond}^n + Nu_{conv}^n)^{(1/n)} \,,
\end{equation}
where $n$ is the closest integer value optimized for the best fit, in this study, we use $n = 5$.  The relationship between the heat transfer($Nu$) and flow conditions ($Ra$) shows that conduction dominates for soils with $Ra < 1$, convection is the dominant mechanism of heat transfer for $Ra > 100$, and a transition region where both conduction and convection heat transfer occurs for $Ra$ between 1 and 100.

\subsection*{Effect of burial depth}
Sediment mobility in submarine conditions may cause a variation in burial depth of up to 5 m in a year~\cite{young2001utilizing}. Changing the cable burial depth alters the distance between the cable and the overlying seawater heat sink, modifying the temperature gradient between the cable and the seawater interface. We vary the burial depth as 0.5, 1.0, and \SI{1.5}{\meter} for three different cable temperatures of \SI{70}{\celsius}, \SI{140}{\celsius}, and \SI{210}{\celsius}. The burial depth and temperature gradient have a linear relation with the Rayleigh-Darcy number. By increasing the burial depth from 0.5 to \SI{1.0}{\meter}, we increase the Rayleigh-Darcy number by a factor of 2. In contrast, by reducing the cable temperature from \SI{140}{\celsius} to \SI{70}{\celsius} the Rayleigh-Darcy number halves. We evaluate the influence of burial depth and cable temperatures by comparing the behavior at the same Rayleigh-Darcy numbers.

\begin{figure}
    \centering
    \includegraphics{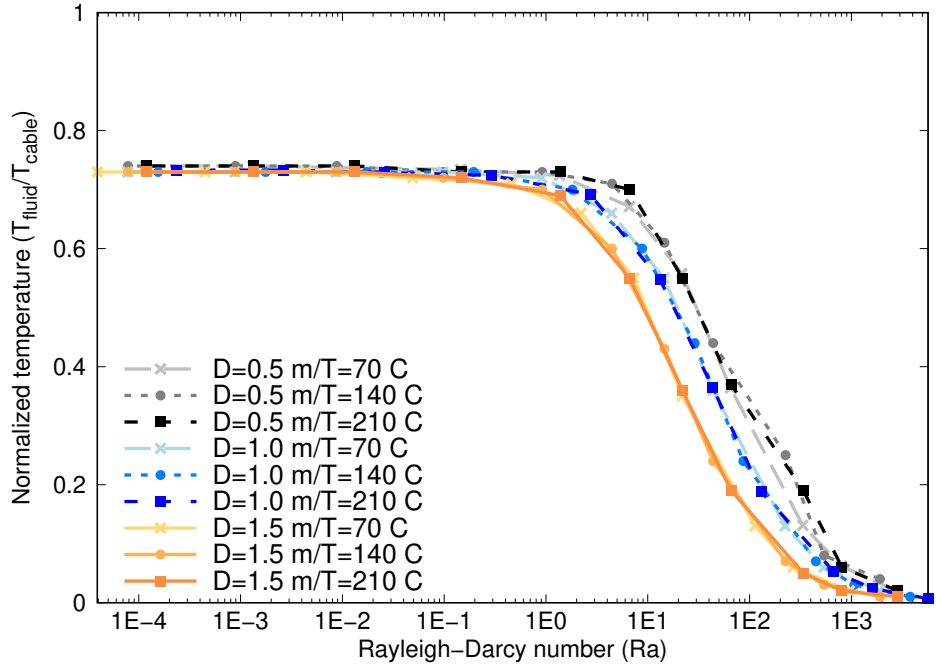}
    \caption{Effect of burial depth on pipe temperature.}
    \label{fig:ra-tnorm-depth}
\end{figure}

As the burial depth increases, the Rayleigh-Darcy number increases, indicating that the deeper the burial is, the more likely the flow is convection.~\Cref{fig:ra-tnorm-depth} shows the heating efficiency for different Rayleigh-Darcy numbers at different burial depths. Heating efficiency is the ratio of pipe temperature to cable temperature.~\Cref{eq:ra} shows that the burial depth influences the time-scale of conduction  ($t \propto D^2$) significantly more than the time-scale of convection ($t \propto 1/D$). In the low permeability condition dominated only by conduction (Ra < 1),~\cref{fig:ra-tnorm-depth} shows the normalized pipe temperature decreasing with the burial depth. However, this decrease in heat transfer efficiency is not significant, dropping by 2\% as burial depth increases from \SI{0.5}{\meter} to \SI{2}{\meter}.

As the Rayleigh-Darcy number increases from 1 to 100, an increase in burial depth shows a marked decrease in the pipeline fluid temperature. In this transition zone, where conduction and convection occur, the increase in burial depth drastically reduces the heat transfer efficiency. The combined effect of quadratic increase in the conduction time scale and the linear drop in the convective time scale deteriorates the heat transfer efficiency. When the permeability becomes more favorable to convective heat transfer ($Ra > 100$), the pipe temperature is almost negligible and remains indifferent to the cable burial depth.

We also observe that the burial depth, not the cable temperature, is the controlling factor in relative heat dissipation. At a given burial depth, the evolution of heating efficiency (normalized pipe fluid temperature) follows the same curve irrespective of the cable temperature. For the range of burial depths, conductive heat transfer efficiency is not affected by the burial depth, while convective heat transfer shows a marked loss in efficiency with an increase in burial depth.

\section{Conclusions}

Deep-Sea pipelines are susceptible to hydrate formation at low temperatures. Inductive heating with piggyback cables reduces this risk and improves the durability of the pipelines by minimizing corrosion. Inductive heating with piggyback cables loses a portion of the generated heat to the surrounding soil. The amount of heat lost to the surrounding environment depends on many parameters: soil thermal conductivity, permeability, cable temperature, and burial depth. The current practice of calculating the current rating required to maintain the cable temperature only considers conductive heat transfer, neglecting the convective heat dissipation. 

We study the different modes of heat transfer for pipelines with inductive heating using 2D Finite Difference analysis. Low permeability soil predominantly experiences conductive heat transfer showing a uniform symmetric heat distribution, thus losing minimum heat to the surrounding soil;  whereas highly permeable soils dissipate a significant portion of the heat to the soil above it, barely heating the pipeline due to buoyancy-driven convective flow driving the heat away from the pipeline and towards the ground surface. A transition region exists where both conduction and convection occur. The Rayleigh-Darcy number, not just permeability, is the controlling factor influencing the transition in the heat transfer behavior. We identify a critical Rayleigh-Darcy number of 1 as the controlling value separating conduction-dominated from convection-dominated heat transfer, much below the established boundary of $Ra = 12$ in the literature for rectangular domains with a constant heat source. At $Ra < 1$, conduction dominates the heat dissipation from the cables and is the ideal zone for inductive heating of pipelines. As Ra increases in the range of 1 - 100, both conduction and convection occur, and heating efficiency significantly drops from 73.5\% to 13\%. When $Ra > 100$, almost no pipeline heating occurs for high permeability soils, losing all the heat to the ground above. For low permeability soil conditions, where only conduction occurs, the depth of burial has a negligible effect on the heating efficiency. Whereas for soils with high permeabilities, an increase in burial depth significantly decreases the heating efficiency as convection dominates.
\pagebreak
\appendix
\section{Notation}
\label{app:notation}
\nopagebreak
\begin{tabular}{r  @{\hspace{1em}=\hspace{1em}}  l}
$\alpha$    & Thermal diffusivity (\si{\meter\squared\per\second})\\
$\beta$     & Thermal expansion of water (\si{\per\kelvin}); \\
$\lambda$   & Thermal conductivity (\si{\watt\per\meter-\kelvin}); \\
$\mu$       & Dynamic viscosity of fluid (\si{\pascal\cdot\second}); \\
$\rho$      & Mass density (\si{\kilogram\per\meter\cubed}); \\
$\nabla$    & Gradient; \\

$c$         & Courant–Friedrichs–Lewy number; \\
$c_{p}$     & Specific heat capacity (\si{\joule\per\kilo\gram\per\celsius}); \\
$d$         & Cable diameter (\si{\meter}); \\
$d_m$       & Mean grain diameter (\si{\meter}); \\
$e$         & Void ratio; \\
$D$         & Depth of burial (\si{\meter}); \\
$g$         & Acceleration due to gravity (\si{\meter\per\second\squared});\\
$G_s$       & Specific gravity of solid; \\
$h$         & Convective heat transfer coefficient (\si{\watt\per\meter\squared\kelvin}); \\
$k$         & Intrinsic permeability (\si{\meter\squared}); \\
$n$         & Porosity; \\
$Nu$        & Nusselt number; \\
$q$         & Heat flux (\si{\watt}); \\
$Ra^\prime$ & Rayleigh number; \\
$Ra$        & Rayleigh-Darcy number; \\
$t$         & time step (\si{\second}); \\
$\Delta x$   & Mesh size (\si{\meter}); \\
$T$         & Temperature (\si{\celsius}); \\
$\mathbf{u}$ & Fluid velocity (\si{\meter\per\second}); \\
\end{tabular}

\subsection{Subscripts}
\begin{tabular}{r  @{\hspace{1em}=\hspace{1em}}  l}
$_0$        & Initial state; \\
$_b$        & Bulk property; \\
$_f$        & Fluid property; \\
$_p$        & Pipe property; \\
$_s$        & Solid property; \\
$_{sat}$    & Saturated property; \\
$_w$        & Water property; \
\end{tabular}

\label{section:references}
\bibliography{ascexmpl-new}

\end{document}